\documentclass{article}
\usepackage{spconf,amsmath,graphicx,hyperref}
\usepackage{amsfonts}
\usepackage{subcaption}
\usepackage{makecell}
\usepackage{tikz}
\usepackage{booktabs}
\usepackage{cite}
\usetikzlibrary{positioning,matrix,arrows.meta,calc,pgfplots.groupplots}
\tikzset{
  token/.style={draw, rounded corners=1pt, fill=gray!10, minimum height=3mm, inner sep=0.8mm},
  good/.style={fill=green!30},
  bad/.style={fill=red!30},
  bar/.style={draw, fill=black!12, minimum height=2.2mm, inner sep=0pt},
  arrow/.style={-Latex, thick},
}

\usepackage{pgfplots}
\usepackage{multirow}
\usepackage{tablefootnote}
\usepackage{enumitem}
\pgfplotsset{compat=1.18}

\newcommand{\eer}{\text{EER}\,\downarrow}
\newcommand{\auc}{\text{AUC}\,\uparrow}
\newcommand{\acc}{\text{ACC}\,\uparrow}

\title{No Word Left Behind: Mitigating Prefix Bias in Open-Vocabulary Keyword Spotting}
\name{Yi Liu$^{1}$\thanks{\textsuperscript{*}Work done during an internship at Bose Corporation.}\textsuperscript{*} \quad
      Chuan-Che (Jeff) Huang$^{2}$ \quad
      Xiao Quan$^{2}$}

\address{$^{1}$ University of California San Diego, Dept. of Electrical and Computer Engineering \\
         $^{2}$ Bose Corporation}

\usepackage[]{todonotes}
\usepackage{amsthm}
\newtheorem{definition}{Definition}[section]
\definecolor{exactblue}{HTML}{4C4CFF}
\definecolor{exactorange}{HTML}{FF9933}
\definecolor{exactgreen}{HTML}{66FF66}
\usepackage[absolute]{textpos}
\begin{document}
\ninept
\maketitle
\begin{textblock*}{18cm}(1.5cm,26.0cm) 
    \noindent\footnotesize\centering
    Copyright 2026 IEEE. Published in ICASSP 2026 - 2026 IEEE International Conference on Acoustics, Speech and Signal Processing (ICASSP), scheduled for 3-8 May 2026 in Barcelona, Spain. Personal use of this material is permitted. However, permission to reprint/republish this material for advertising or promotional purposes or for creating new collective works for resale or redistribution to servers or lists, or to reuse any copyrighted component of this work in other works, must be obtained from the IEEE. Contact: Manager, Copyrights and Permissions / IEEE Service Center / 445 Hoes Lane / P.O. Box 1331 / Piscataway, NJ 08855-1331, USA. Telephone: + Intl. 908-562-3966.
\end{textblock*}
\begin{abstract}

Open-vocabulary keyword spotting (OV-KWS) enables personalized device control via arbitrary voice commands. Recently, researchers have explored using audio-text joint embeddings, allowing users to enroll phrases with text, and proposed techniques to disambiguate similar utterances. We find that existing OV-KWS solutions often overly bias the beginning phonemes of an enrollment, causing false triggers when negative enrollment-query-pairs share a prefix (``turn the volume up'' vs. ``turn the volume down''). We trace this to two factors: training data bias and position-biased cross-modal scoring. To address these limitations, we introduce the Partial Overlap Benchmark (POB) with two datasets, POB-Spark and POB-LibriPhrase (POB-LP), containing mismatched audio-text pairs with shared prefixes, and propose Equal-weighting Position Scoring (EPS), a lightweight decision layer. Using EPS alone reduces EER on POB-Spark from 64.4\% to 29.3\% and improves POB-LP accuracy from 87.6\% to 96.8\%, while maintaining performance on LibriPhrase and Google Speech Commands (GSC). With POB data added in training, our work achieves the best POB benchmark results while incurring the least amount of degradation on prior metrics among baselines. This degradation is most pronounced in GSC, which contains only one-word commands. We surface mitigating this trade-off as future work.

\end{abstract}


\section{Introduction}
\label{sec:intro}
Open-vocabulary keyword spotting (OV-KWS) can enable many useful applications on edge devices, such as allowing users to define custom shortcuts for their devices (e.g., ``turn the light on'', ``open the garage'') and activating actions in portable game consoles (e.g., say ``thunderbolt'' to trigger a move in a game). 
Unlike keyword spotting with a pre-defined set of commands, OV-KWS allows arbitrary phrases to be enrolled and detected, broadening their utility in controlling edge devices.

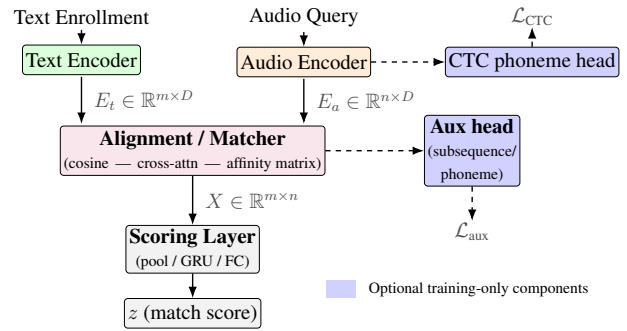
\begin{figure}[t!]
\centering
\resizebox{0.95\columnwidth}{!}{%
\begin{tikzpicture}[>=Latex, node distance=6mm and 10mm, font=\large]
  \tikzset{
    module/.style      ={draw, rounded corners=2pt, fill=black!5, inner sep=2pt, minimum height=6mm, minimum width=20mm, align=center},
    moduleaudio/.style ={module, fill=orange!15},    
    moduletext/.style  ={module, fill=green!14},     
    alignbox/.style    ={module, fill=purple!10},    
    trainonly/.style   ={module, fill=blue!18},      
    labelbox/.style    ={fill=white, inner sep=1pt, text=black!70, font=\large}
  }

  \node (text_in) {Text Enrollment};
  \node[right=18mm of text_in] (audio_in) {Audio Query};

  \node[moduletext,  below=3mm of text_in]  (text_enc)  {Text Encoder};
  \node[moduleaudio, below=3mm of audio_in] (audio_enc) {Audio Encoder};

  \node (mid) at ($(text_enc)!0.5!(audio_enc)$) {};
  \node[alignbox, below=12mm of mid, minimum height=10mm, minimum width=48mm] (align)
        {\textbf{Alignment / Matcher}\\{\normalsize (cosine \,|\, cross-attn \,|\, affinity matrix)}};

  \coordinate (align_in_t) at (align.north -| text_enc.south);
  \coordinate (align_in_a) at (align.north -| audio_enc.south);

  \node[module, below=10mm of align] (agg) {\textbf{Scoring Layer}\\{\normalsize (pool / GRU / FC)}};
  \node[module, below=5mm of agg, minimum width=16mm] (outz) {$z$ (match score)};

  \node[trainonly, right=15mm of audio_enc, minimum width=26mm] (ctc_head) {CTC phoneme head};
  \node[trainonly, right=20mm of align, minimum width=18mm] (aux_head)
        {\textbf{Aux head}\\{\normalsize (subsequence/}\\{\normalsize phoneme)}};

  \draw[->, thick] (text_in)  -- (text_enc);
  \draw[->, thick] (audio_in) -- (audio_enc);

  \draw[->, thick, shorten >=1pt, shorten <=1pt]
    (text_enc.south) -- node[midway, right=6pt, labelbox] {$E_t \in \mathbb{R}^{m \times D}$} (align_in_t);
  \draw[->, thick, shorten >=1pt, shorten <=1pt]
    (audio_enc.south) -- node[midway, right=6pt, labelbox] {$E_a \in \mathbb{R}^{n \times D}$} (align_in_a);

  \draw[->, thick] (align.south) -- node[midway, right=6pt, labelbox] {$X \in \mathbb{R}^{m \times n}$} (agg.north);
  \draw[->, thick] (agg.south) -- (outz.north);

  \draw[->, thick, dashed] (audio_enc.east) -- (ctc_head.west);
  \draw[->, thick, dashed] (align.east)     -- (aux_head.west);

  \node[labelbox, above=4mm of ctc_head] (ctc_loss) {$\mathcal{L}_{\text{CTC}}$};
  \draw[->, thick, dashed] (ctc_head.north) -- (ctc_loss.south);

  \node[labelbox, below=7mm of aux_head] (aux_loss) {$\mathcal{L}_{\text{aux}}$};
  \draw[->, thick, dashed] (aux_head.south) -- (aux_loss.north);

  \node[draw=none, fill=blue!18, minimum width=6mm, minimum height=3.5mm, above=0mm of outz, xshift=30mm] (leg) {};
  \node[anchor=west] at ($(leg.east)+(1.5mm,0)$) {\normalsize Optional training-only components};
\end{tikzpicture}
}
\caption{General OV-KWS system with optional training-only heads.}

\label{fig:general_arch_vertical}
\vspace{-2pt}
\end{figure}

Recent research has explored the use of joint audio-text embeddings for OV-KWS to simplify the enrollment process. A central challenge in this paradigm is aligning the two modalities. Most approaches adopt a three-stage neural architecture: 1) project audio query and enrolled text into a joint embedding space, 2) compute an alignment vector or matrix between the embeddings, and 3) apply a scoring layer to determine a matching score, as shown in \autoref{fig:general_arch_vertical}. 

To bridge the gap between audio and text modalities, many studies introduce auxiliary objectives during training to improve embedding alignment. For instance, phoneme-level objectives have been applied in both embedding alignment and audio encoder training \cite{nishu2025slick,lee2023phonmatchnet,nishu2024flexible}. A range of alignment architectures have also been explored, including self-attention \cite{lee2023phonmatchnet,li2025phoneme,kim2024bridging}, cross-attention\cite{nishu2025slick,li2025phoneme,kim2024bridging,jung25b_interspeech}, dynamic sequence partitioning\cite{nishu2023matching,nishu2024flexible}, and CTC forced alignment\cite{kim2025fully}. In addition, generative models have been leveraged to augment training data by synthesizing hard negatives––pairs of audio and text with similar phonetic content but different meanings \cite{zhu2024ge2e,zhang2025graphemeaug,zhu25b_interspeech}.

Among prior work, PhonMatchNet\cite{lee2023phonmatchnet} and SLiCK\cite{nishu2025slick} are notable examples that achieve state-of-the-art performance on the LibriPhrase dataset while using less than one million parameters during inference, making them feasible for edge deployment. To achieve this result, PhonMatchNet uses self-attention as the alignment mechanism to compare the audio and text embeddings at both the phoneme and utterance level. SLiCK uses a similar design but utilizes cross-attention as the alignment module. Additionally, it imposes a length constraint of 25 phonemes on the cross-attention layer to reduce computation complexity.

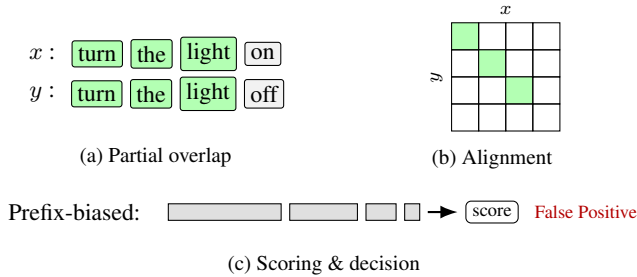
\begin{figure}[htbp]
\centering

\begin{minipage}[c]{0.48\columnwidth}
\centering
\vspace{5mm}
\begin{tikzpicture}[node distance=1mm]
  \node (xl) {$x$ :};
  \node[token, good, right=1mm of xl] (x1) {turn};
  \node[token, good, right=1mm of x1] (x2) {the};
  \node[token, good, right=1mm of x2] (x3) {light};
  \node[token, right=1mm of x3] (x4) {on};
  \node (yl) [below=1.3mm of xl] {$y$ :};
  \node[token, good, right=1mm of yl] (y1) {turn};
  \node[token, good, right=1mm of y1] (y2) {the};
  \node[token, good, right=1mm of y2] (y3) {light};
  \node[token, right=1mm of y3] (y4) {off};
\end{tikzpicture}
\vspace{2.5mm}
\subcaption{Partial overlap}
\end{minipage}\hfill
\begin{minipage}[c]{0.48\columnwidth}
\centering
\begin{tikzpicture}[node distance=1mm]
  \newcommand{\cell}{3.6mm}

  \node (ry1) {};                       
  \node (ry2) [below=3.6mm of ry1] {};
  \node (ry3) [below=3.6mm of ry2] {};
  \node (ry4) [below=3.6mm of ry3] {};

  \node (col0) [right=0mm of ry2] {};

  \node[draw, minimum width=\cell, minimum height=\cell, fill=green!30, anchor=west]
    (r1c1) at ($(col0.east |- ry1.center)$) {};
  \node[draw, minimum width=\cell, minimum height=\cell, right=-\pgflinewidth of r1c1] (r1c2) {};
  \node[draw, minimum width=\cell, minimum height=\cell, right=-\pgflinewidth of r1c2] (r1c3) {};
  \node[draw, minimum width=\cell, minimum height=\cell, right=-\pgflinewidth of r1c3] (r1c4) {};

  \node[draw, minimum width=\cell, minimum height=\cell, below=-\pgflinewidth of r1c1] (r2c1) {};
  \node[draw, minimum width=\cell, minimum height=\cell, fill=green!30, right=-\pgflinewidth of r2c1] (r2c2) {};
  \node[draw, minimum width=\cell, minimum height=\cell, right=-\pgflinewidth of r2c2] (r2c3) {};
  \node[draw, minimum width=\cell, minimum height=\cell, right=-\pgflinewidth of r2c3] (r2c4) {};

  \node[draw, minimum width=\cell, minimum height=\cell, below=-\pgflinewidth of r2c1] (r3c1) {};
  \node[draw, minimum width=\cell, minimum height=\cell, right=-\pgflinewidth of r3c1] (r3c2) {};
  \node[draw, minimum width=\cell, minimum height=\cell, fill=green!30, right=-\pgflinewidth of r3c2] (r3c3) {};
  \node[draw, minimum width=\cell, minimum height=\cell, right=-\pgflinewidth of r3c3] (r3c4) {};

  \node[draw, minimum width=\cell, minimum height=\cell, below=-\pgflinewidth of r3c1] (r4c1) {};
  \node[draw, minimum width=\cell, minimum height=\cell, right=-\pgflinewidth of r4c1] (r4c2) {};
  \node[draw, minimum width=\cell, minimum height=\cell, right=-\pgflinewidth of r4c2] (r4c3) {};
  \node[draw, minimum width=\cell, minimum height=\cell, right=-\pgflinewidth of r4c3] (r4c4) {};

  \node[anchor=south] at ($(r1c1.north)!0.5!(r1c4.north)$) {\scriptsize $x$};
  \node[rotate=90, anchor=south] at ($(r1c1.west)!0.5!(r4c1.west)$) {\scriptsize $y$};
\end{tikzpicture}
\vspace{-6mm}
\subcaption{Alignment}
\end{minipage}

\par\vspace{3mm}
\begin{minipage}[t]{\columnwidth}
\begin{tikzpicture}[node distance=2.2mm]
  \node (l1) {Prefix-biased:};
  \node[bar, minimum width=15mm, right=2.5mm of l1] (pb1) {};
  \node[bar, minimum width=9mm,  right=1mm  of pb1] (pb2) {};
  \node[bar, minimum width=4mm,  right=1mm  of pb2] (pb3) {};
  \node[bar, minimum width=2mm,  right=1mm  of pb3] (pb4) {};
  \node[draw, rounded corners=2pt, inner sep=2.5pt, font=\scriptsize, right=6.0mm of pb4] (s1) {score};
  \draw[arrow, shorten >=1mm, shorten <=1mm] (pb4.east) -- (s1.west);
  \node[text=red!70!black, right=1mm of s1] {\scriptsize False Positive};

  \node (gap) [below=-1mm of l1] {};


\end{tikzpicture}
\subcaption{Scoring \& decision}
\end{minipage}
\caption{Partial-overlap failure: query “turn the light off” vs. enrollment “turn the light on”. Top: (a) tokens and (b) alignment. Bottom: (c) a scorer that overweights early positions, producing a false positive }


\label{fig:partial-overlap-illustration}
\end{figure}

Although these works have achieved impressive performance on LibriPhrase test sets, we find that their performance drops sharply once the enrolled text contains three or more words (10 or more phonemes) and when they share an overlapping prefix as the query. We show an example of such error case in \autoref{fig:partial-overlap-illustration}. Our experiments point to two causes. First, the training and test sets used in these work (e.g., LibriPhrase and Google Speech Commands) contain few phrases with overlapping prefixes and are skewed toward phrases that are two words or shorter, leaving longer overlapping cases underrepresented. Second, existing scoring layers tend to easily bias toward early phonemes, further amplifying prefix confusion. As a result, both current models and benchmarks fail to address the ambiguity posed by partially overlapping query–enrollment pairs.

In this work, we improve OV-KWS by making the following contributions: 
\begin{enumerate}
    \item  We show that existing OV-KWS solutions fail to handle partially overlapping phrases, especially phrases sharing two or more common prefix words.
    \item  We then create two datasets, one derived from LibriPhrase and the other synthesized using a state-of-the-art text-to-speech model. These datasets form the basis of our Partial Overlap Benchmark to improve and better capture performance of OV-KWS in practical applications. \footnote{We release our dataset at \url{https://github.com/cijinsama/Partial-Overlap-Benchmark}}
    \item  Finally, we design a lightweight equal-weighting position scoring (EPS) module and demonstrate that with EPS, we can achieve a 35.1$\%$ reduction in EER (64.4$\%$ $\rightarrow$ 29.3$\%$) on POB-Spark and a 9.2$\%$ improvement in accuracy (87.6$\%$ $\rightarrow$ 96.8$\%$) on POB-LP over previous work in differentiating longer phrases with overlapping prefix without sacrificing accuracy and computational efficiency on prior benchmarks. 
\end{enumerate}


\section{Partial Overlap Benchmark}
\label{sec:pob}

Existing benchmarks rarely evaluate audio queries that match only a portion of an enrolled phrase, resulting in insufficient coverage for cases where enrolled utterances contain multiple words with overlapping prefixes. In this section, we first define \emph{partial overlap} and \emph{first-different phoneme index}, then apply these concepts to analyze LibriPhrase, a dataset commonly used for OV-KWS evaluation. Finally, we introduce our Partial Overlap Benchmark (POB) to address this gap. 

\subsection{Partial Overlap}
We define \emph{partial overlap} as a query that matches only the beginning part of an enrolled phrase. Let the enrollment be a token sequence
$\mathbf{x} = (x_1, \dots, x_p)$ and the token sequence corresponding to the query audio be $\mathbf{y} = (y_1, \dots, y_q)$. 

\begin{definition}[Partial Overlap]
A pair $(\mathbf{x}, \mathbf{y})$ exhibits partial overlap if
\[
\bigl(y_i = x_i \;\; \forall i=1,\dots,q \bigr) \;\wedge\; (p > q).
\]
\end{definition}

Thus $\mathbf{y}$ coincides with the prefix of $\mathbf{x}$ but differs at later positions.

\subsection{First-different Phoneme Index}
We further define the \emph{first-different phoneme index} as a measure of the degree of partial overlap in a dataset. For any two words, this index denotes the position of the first phoneme at which they differ. Using the first-different phoneme index to analyze LibriPhrase (see \autoref{fig:data-distribution}), we find that the dataset primarily contains utterances sharing the first 0-4 phonemes. This limitation makes LibriPhrase less suitable for evaluating performance in applications where enrollments may consist of multiple words and common prefixes.

\subsection{Design of Partial Overlap Benchmark}
To address the aforementioned limitation, we introduce \emph{Partial Overlap Benchmark (POB)}. POB contains two datasets: 

\begin{itemize} [leftmargin=1em]
    \item \emph{POB-LP} A LibriPhrase-derived dataset created by appending an additional word from 10,000 most common English words \cite{google-10000-english} to the enrollment text sequence to simulate prefix overlaps. POB-LP is designed to follow the same phrase length distribution as the original LibriPhrase. 
    \item \emph{POB-Spark} A synthetic dataset generated with a state-of-the-art TTS model, Spark-TTS \cite{wang2025sparktts}, to provide controlled overlap patterns across varied speaker characteristics. Compared to POB-LP, POB-Spark contains a balanced distribution across phrase lengths. We constructed this dataset with the following steps:
    \begin{enumerate}[leftmargin=1.5em]
    \item \textbf{Word–phoneme mapping.} Each word $w \in W$ from common English words\cite{google-10000-english} is mapped to its CMU phoneme sequence $P(w)$, with length $L(w)=|P(w)|$.
    
    \item \textbf{Closest match.} For each $w$, find its phonetic neighbors. Phoneme similarity is measured by Levenshtein distance $\delta(\cdot,\cdot)$.
    
    \item \textbf{Phrase construction.} Randomly sample words $\{w_i\}$ such that $\sum_i L(w_i)<L_{\max}$, yielding a phrase.
    
    \item \textbf{Pair construction.} Replace one randomly chosen word $w_j$ from a phrase with it's phonetic neighbors $w'_j$, yielding a phrase pair.
    
    \item \textbf{Sample.} From all candidate pairs, compute real pronunciations by G2P\cite{g2pE2019} and the first differing index $i_\text{diff}$. Then sample such that $i_\text{diff}$ is approximately uniformly distributed.
    
    \item \textbf{Final dataset.} Each pair $(a,b)$ yields $(a,b,\text{False})$, \\$(b, a,\text{False})$ and $(a,a,\text{True})$ query-anchor-label tuples.
    \end{enumerate}

\end{itemize}

\begin{figure}[t]
\centering


\begin{subfigure}{\columnwidth}
  \centering
\begin{tikzpicture}
\begin{axis}[
    ybar,
    bar width=9pt,
    width=\linewidth,
    height=3.5cm,
    enlarge x limits=0.12,
    ylabel=Ratio,
    xlabel={First Different Phoneme Index},
    symbolic x coords={{[0,4]},{[5,9]},{[10,14]},{[15,19]},{[20,24]}},
    xtick=data,
    ymin=0, ymax=1.0,
    ymajorgrids=true,
    legend style={at={(0.97,0.97)},anchor=north east, font=\scriptsize, fill=white},
    legend cell align=left,
    cycle list={{fill=blue!45,fill opacity=1.0},
                {fill=orange!45,fill opacity=1.0},
                {fill=green!45,fill opacity=1.0}},
    every axis plot/.append style={line width=0.25pt},
]

\addplot coordinates {
    ({[0,4]},0.9309636534099972) ({[5,9]},0.06672553909116019) ({[10,14]},0.002217441539293453) ({[15,19]},9.336595954919803e-05) ({[20,24]},0.0)
};
\addplot coordinates {
    ({[0,4]},0.13231241144138392) ({[5,9]},0.7485168864736932) ({[10,14]},0.11411799511078585) ({[15,19]},0.005013780109929465) ({[20,24]},3.89268642075269e-05)
};
\addplot coordinates {
    ({[0,4]},0.20911458333333333) ({[5,9]},0.20828125) ({[10,14]},0.2090625) ({[15,19]},0.24046875) ({[20,24]},0.13307291666666668)
};

\legend{LibriPhrase, POB-LP, POB-Spark}

\end{axis}
\end{tikzpicture}
\end{subfigure}

  \caption{Distribution of first-different phoneme index.  
  LibriPhrase diverges almost immediately, whereas POB-LP and
  POB-Spark contain longer common prefixes before mismatch occurs, enabling more diverse overlapping cases for evaluation.}
\label{fig:data-distribution}
\vspace{-4mm}
\end{figure}
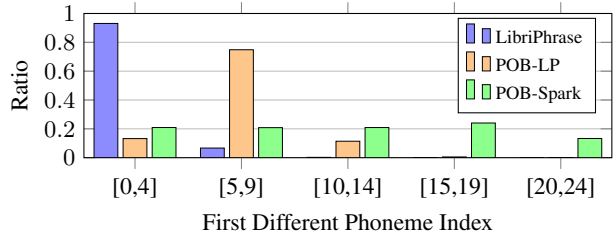

As shown in \autoref{fig:data-distribution}, the POB datasets contain far more overlapping prefixes across all phrase lengths than the original LibriPhrase, enabling more realistic and challenging evaluation.

\section{Prefix Bias and EPS Module}


In addition to positional skews in the training data, we also find that architectural design plays an important role in amplifying such biases. In this section, we first define \emph{Prefix Bias}, highlight its existence in a state-of-the-art work, and present an alternative scoring design aimed at alleviating symptoms caused by such biases.

\subsection{Prefix Bias}
As shown in \autoref{fig:general_arch_vertical}, we denote the hidden variable after the Alignment/Matcher module as $X \in \mathbb{R}^{m\times n}$. $X_i$ encodes the degree to which the anchor text phoneme at position $i$ matches to the query audio. Subsequently, a scoring module assigns position-dependent weights $a_i$ to the vectors and aggregates them to produce the final binary decision $z$. We see this pattern in both SLiCK and PhonMatchNet, two sub-1M-parameter models mentioned in \autoref{sec:intro}. In SLiCK, this is done by utterance-level matching FC layer and softmax, and in PhonMatchNet, a GRU paired with an FC layer and sigmoid activation.
\begin{definition}[Prefix Bias]
Let $X = [X_1,\dots,X_m]\in\mathbb{R}^{m\times n}$ and
$A = [a_1,\dots,a_m]$. The expected contribution of position $i$ is
\[
C_i = \mathbb{E}_{X_i}[|a_i^\top X_i|].
\]

We define the \emph{\textbf{prefix concentration score}}
\[
\rho(k) = \frac{\sum_{i=1}^k \|a_i\|_2}{\sum_{i=1}^m \|a_i\|_2},
\]
which measures how much scoring weight is concentrated in the first
$k$ positions.
\end{definition}


\subsection{Prefix Bias Can be Found in Prior Work}
Our experiments indicate that prefix bias can be directly observed in previous work through matcher module weight visualization. Using SLiCK as an example, ~\autoref{fig:prefix-diagnostics} visualizes the position-wise contributions $C_i$ of the matcher module: the SLiCK baseline over-weighs earlier phoneme indices much more than later ones. This suggests that the trained module will prioritize matching earlier sections of an audio-text pair, potentially ignoring later mismatches. This occurs even when subsequence-level matching loss is used -- one of the main contributions of SLiCK -- as an auxiliary task during training, designed to disambiguate similar sounding audio-text pairs with common phoneme components, e.g. ``blue'' vs ``glue''. In the next section we introduce our alternative scorer module design that specifically addresses this bias.


\subsection{Equal-weighting Position Scoring (EPS) Module}
To mitigate prefix bias, we propose \emph{Equal-weighting Position Scoring} module, a lightweight change applied only at the final scoring stage of OV-KWS models. Instead of flattening time-aligned embeddings into a position-biasing FC layer, we replace it with a single position-independent linear map followed by average pooling, see \autoref{fig:arch}:
\[
z_i = w^\top X_i, \qquad 
z = \tfrac{1}{m}\sum_{i=1}^m z_i + b,
\]
Here, $\mathbf{X}\!\in\!\mathbb{R}^{m\times n}$ denotes the sequence of alignment vectors
from the audio–text matcher, where $m$ is the number of token positions and $n$ the
per-position feature dimension. All positions
share the same weight vector $w$, so contributions are uniform across time. This modification adds no additional parameters and removes dependence on a fixed input-sequence-length before the matcher module. We position EPS as a minimal, interpretable baseline; future work could explore constrained dynamic attention or positional weight regularization to improve discriminative power without re-introducing bias.


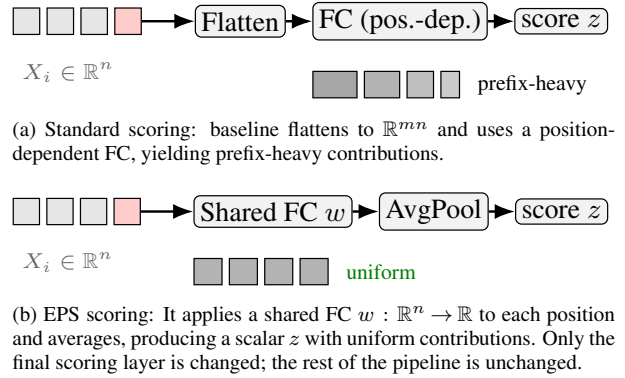
\begin{figure}[t]
\centering

\begin{subfigure}[b]{0.45\textwidth}
\centering
\resizebox{\textwidth}{!}{%
\begin{tikzpicture}[node distance=2.5mm, >=Latex]
  \tikzset{
    vec/.style    ={draw, minimum width=3mm, minimum height=3mm, fill=black!10},
    module/.style ={draw, rounded corners=2pt, fill=black!5, inner sep=2pt},
    dim/.style    ={font=\scriptsize, text=black!60}
  }

  \node[vec] (x1) {};
  \node[vec, right=0.7mm of x1] (x2) {};
  \node[vec, right=0.7mm of x2] (x3) {};
  \node[vec, right=0.7mm of x3, fill=red!20] (x4) {};
  \node[anchor=north west, font=\scriptsize, text=black!60]
    at ($(x1.south west)+(0,-2.0mm)$) {$X_i \in \mathbb{R}^{n}$};

  \node[module, right=6mm of x4] (flat) {Flatten};
  \node[module, right=3mm of flat] (fcdep) {FC (pos.-dep.)};
  \node[module, right=3mm of fcdep] (score1) {score $z$};

  \draw[arrow] (x4.east) -- (flat.west);
  \draw[arrow] (flat.east) -- (fcdep.west);
  \draw[arrow] (fcdep.east) -- (score1.west);


  \node[vec, fill=black!35, below=5.0mm of fcdep.south west, anchor=west, minimum width=5mm] (b1) {};
  \node[vec, fill=black!30, right=0.7mm of b1, minimum width=4mm] (b2) {};
  \node[vec, fill=black!25, right=0.7mm of b2, minimum width=3mm] (b3) {};
  \node[vec, fill=black!20, right=0.7mm of b3, minimum width=2mm] (b4) {};
  \node[anchor=west, font=\scriptsize] at ($(b4.east)+(0.9mm,0)$) {prefix-heavy};
\end{tikzpicture}
}
\subcaption{Standard scoring: baseline flattens to
$\mathbb{R}^{mn}$ and uses a position-dependent FC, yielding prefix-heavy
contributions.}
\end{subfigure}

\vspace{3mm}

\begin{subfigure}[b]{0.45\textwidth}
\centering
\resizebox{\textwidth}{!}{%
\begin{tikzpicture}[node distance=2.5mm, >=Latex]
  \tikzset{
    vec/.style    ={draw, minimum width=3mm, minimum height=3mm, fill=black!10},
    module/.style ={draw, rounded corners=2pt, fill=black!5, inner sep=2pt},
    dim/.style    ={font=\scriptsize, text=black!60}
  }

  \node[vec] (y1) {};
  \node[vec, right=0.7mm of y1] (y2) {};
  \node[vec, right=0.7mm of y2] (y3) {};
  \node[vec, right=0.7mm of y3, fill=red!20] (y4) {};
  \node[anchor=north west, font=\scriptsize, text=black!60]
    at ($(y1.south west)+(0,-2.0mm)$) {$X_i \in \mathbb{R}^{n}$};

  \node[module, right=6mm of y4] (shared) {Shared FC $w$};
  \node[module, right=3mm of shared] (pool) {AvgPool};
  \node[module, right=3mm of pool] (score2) {score $z$};

  \draw[arrow] (y4.east) -- (shared.west);
  \draw[arrow] (shared.east) -- (pool.west);
  \draw[arrow] (pool.east) -- (score2.west);


  \node[vec, fill=black!30, below=5.0mm of shared.south west, anchor=west, minimum width=3.2mm] (e1) {};
  \node[vec, fill=black!30, right=0.7mm of e1, minimum width=3.2mm] (e2) {};
  \node[vec, fill=black!30, right=0.7mm of e2, minimum width=3.2mm] (e3) {};
  \node[vec, fill=black!30, right=0.7mm of e3, minimum width=3.2mm] (e4) {};
  \node[anchor=west, font=\scriptsize, text=green!50!black]
    at ($(e4.east)+(0.9mm,0)$) {uniform};
\end{tikzpicture}
}
\subcaption{EPS scoring: It applies a shared FC
$w:\mathbb{R}^{n}\!\to\!\mathbb{R}$ to each position and averages, producing a
scalar $z$ with uniform contributions. Only the final scoring layer is changed;
the rest of the pipeline is unchanged.}
\end{subfigure}

\vspace{-1mm}
\caption{Final scoring stage in SLiCK. Gray squares denote aligned per-position
features ($X_i \!\in\! \mathbb{R}^{n}$) from the preceding alignment module,
stacked as $X \!\in\! \mathbb{R}^{m\times n}$.}
\label{fig:arch}
\vspace{-4mm}
\end{figure}
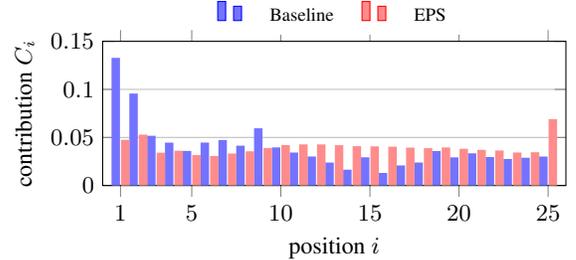
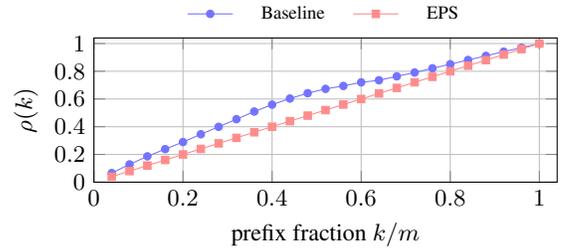
\begin{figure}[htbp]
\vspace{-4pt}
\centering

\begin{subfigure}[b]{0.9\columnwidth}
\centering
\begin{tikzpicture}
\begin{axis}[
  ybar, bar width=4pt,
  yticklabel style={/pgf/number format/fixed},
  scaled y ticks=false,
  ymin=0, ymax=0.15,
  xmin=0, xmax=26,
  width=\columnwidth,
  height=3.5cm,
  xlabel={position $i$}, ylabel={contribution $C_i$},
  xtick={1,5,10,15,20,25},
  ymajorgrids,
  legend style={font=\scriptsize, at={(0.5,1.32)}, anchor=north, draw=none, legend columns=2, column sep=3mm},
]
  \addplot+[ybar, bar width=3.2pt, bar shift=-1.8pt, fill=blue!55, draw=none] coordinates {
    (1,0.13278139) (2,0.09565169) (3,0.05172081) (4,0.04459627) (5,0.03595852)
    (6,0.04468333) (7,0.0472786)  (8,0.04142889) (9,0.05951905) (10,0.03961303)
    (11,0.03427947) (12,0.03017307) (13,0.02387768) (14,0.01643184) (15,0.02933976)
    (16,0.01307582) (17,0.02087198) (18,0.02390878) (19,0.03585902) (20,0.02925892)
    (21,0.03347726) (22,0.02960924) (23,0.02759231) (24,0.02889202) (25,0.03012124)
  };
  \addplot+[ybar, bar width=3.2pt, bar shift=+1.8pt, fill=red!45, draw=none] coordinates {
    (1,0.04753224) (2,0.05291966) (3,0.03413335) (4,0.03621528) (5,0.0316368)
    (6,0.030677)   (7,0.03329291) (8,0.03558628) (9,0.03892606) (10,0.04208338)
    (11,0.04277199) (12,0.04275058) (13,0.04196191) (14,0.04091186) (15,0.04070839)
    (16,0.04034415) (17,0.03938742) (18,0.03885694) (19,0.03961523) (20,0.03819575)
    (21,0.03706873) (22,0.03643133) (23,0.03432996) (24,0.03461402) (25,0.06904879)
  };
  
  \legend{Baseline, EPS}
\end{axis}
\end{tikzpicture}
\subcaption{Position-wise contributions $C_i$ (normalized): the baseline concentrates weight on early
positions (tall left bars), whereas EPS distributes weight more uniformly.}
\end{subfigure}

\par\vspace{2mm}

\begin{subfigure}[b]{0.9\columnwidth}
\centering
\begin{tikzpicture}
\begin{axis}[
  xmin=0, xmax=1.04, ymin=0, ymax=1.04,
  width=\columnwidth,
  height=3.5cm,
  xlabel={prefix fraction $k/m$}, ylabel={$\rho(k)$},
  grid=both,
  legend style={font=\scriptsize, at={(0.5,1.3)}, anchor=north, draw=none, legend columns=2, column sep=3mm},
]
    \addplot+[blue!55, mark size=1.4pt, mark options={solid}] coordinates {
      (0.04,0.065676) (0.08,0.128425) (0.12,0.186410) (0.16,0.238341) (0.20,0.289423)
      (0.24,0.345606) (0.28,0.399385) (0.32,0.454075) (0.36,0.509438) (0.40,0.559142)
      (0.44,0.603235) (0.48,0.641839) (0.52,0.672874) (0.56,0.693918) (0.60,0.720074)
      (0.64,0.735084) (0.68,0.763347) (0.72,0.790770) (0.76,0.821348) (0.80,0.850867)
      (0.84,0.881231) (0.88,0.911012) (0.92,0.941171) (0.96,0.970116) (1.00,1.000000)
    };
  \addplot+[red!45, mark=square*, mark size=1.4pt, mark options={solid}] coordinates {
    (0.04,0.04) (0.08,0.08) (0.12,0.12) (0.16,0.16) (0.20,0.20) (0.24,0.24)
    (0.28,0.28) (0.32,0.32) (0.36,0.36) (0.40,0.40) (0.44,0.44) (0.48,0.48)
    (0.52,0.52) (0.56,0.56) (0.60,0.60) (0.64,0.64) (0.68,0.68) (0.72,0.72)
    (0.76,0.76) (0.80,0.80) (0.84,0.84) (0.88,0.88) (0.92,0.92) (0.96,0.96)
    (1.00,1.00)
  };
  
  \legend{Baseline, EPS}
\end{axis}
\end{tikzpicture}
\subcaption{Cumulative prefix concentration $\rho(k)$ vs. prefix fraction $k/m$: the baseline
rises steeply above the identity (high prefix concentration), while EPS
tracks the identity (little to no prefix bias).}
\end{subfigure}

\vspace{-1mm}
\caption{Prefix-bias diagnostics with $m{=}25$ positions}
\label{fig:prefix-diagnostics}
\vspace{-1mm}
\end{figure}

\section{Experiments}
\label{sec:experiments}

\begin{table*}[t]
\centering
\caption{Results across two training regimes. 1. \textbf{LibriPhrase-only}: isolates the architectural effects of using EPS. 2. \textbf{POB data augmentation}: incorporates additional POB data during training. This regime highlights training data composition effects.}

\label{fig:results-pack}
\vspace{-2mm}
\begingroup
\small
\setlength{\tabcolsep}{4.2pt}
\renewcommand{\arraystretch}{0.95}

\begin{tabular}{l r cc cc cc cc c}
\toprule
\multirow{2}{*}{Model} & \multirow{2}{*}{Param} &
\multicolumn{2}{c}{LP-easy} & \multicolumn{2}{c}{LP-hard} &
\multicolumn{2}{c}{GSC} & \multicolumn{2}{c}{POB-Spark} & POB-LP \\
\cmidrule(lr){3-4}\cmidrule(lr){5-6}\cmidrule(lr){7-8}\cmidrule(lr){9-10}
 & & $\eer$ & $\auc$ & $\eer$ & $\auc$ & $\eer$ & $\auc$ & $\eer$ & $\auc$ & $\acc$ \\
\midrule
PhonMatchNet\footnotemark[1] & 655K & 4.49\% & 99.02\% & 23.18\% & 84.55\% & 10.28\% & 95.95\% & 34.98\% & 70.14\% & 64.30\% \\
SLiCK\footnotemark[2]        & 580K & 2.14\% & 99.76\% & 14.30\% & 91.77\% & \textbf{8.00}\%  & \textbf{97.52}\% & 64.41\% & 31.34\% & 87.62\% \\
\textbf{SLiCK-EPS (ours)}    & 557K & \textbf{1.82\%} & \textbf{99.80\%} & \textbf{13.70\%} & \textbf{92.66\%} & 8.87\% & 97.19\% & \textbf{29.28\%} & \textbf{77.47\%} & \textbf{96.82\%} \\
\addlinespace[2pt]
\midrule
PhonMatchNet + POB training  &   --  & 11.72\% & 95.25\% & 29.81\% & 76.77\% & 27.73\% & 80.52\% & 18.68\% & 89.25\% & \textbf{99.87\%} \\
SLiCK + POB training         &   --  & 5.26\%  & 98.69\% & 25.46\% & 81.52\% & 25.92\% & 80.92\% & 29.23\% & 77.88\% & 98.70\% \\
\textbf{SLiCK-EPS + POB training (ours)} & -- & \textbf{3.24\%} & \textbf{99.49\%} &\textbf{ 17.75\%} & \textbf{89.41\%} &
\textbf{18.75\%} &\textbf{89.31\%} & \textbf{16.15\%} & \textbf{91.14\%} & 99.42\% \\
\bottomrule
\end{tabular}
\endgroup
\end{table*}

To evaluate the effect of adding POB data and applying our EPS module at the final output layer, we conduct controlled experiments across multiple training conditions. Specifically, we modify only the final scoring layer of SLiCK (\autoref{fig:arch}) to use the proposed EPS module, and refer to this variant as \textbf{SLiCK-EPS}. This isolates the scoring mechanism while keeping the encoder and alignment modules unchanged, enabling us to directly attribute performance differences to the scoring design.

\emph{Datasets:} We first use LibriPhrase\cite{shin2022learning} as training dataset (182,570 samples). Then combine it with our proposed POB training set (90,808 samples) for comparison. For evaluation, we use the Libriphrase-hard, Libriphrase-easy, Google Speech Commands\cite{speechcommandsv2} and our proposed two subsets: POB-Spark and POB-LP as test set.

\emph{Baselines:} At time of writing, we evaluate SLiCK and PhonMatchNet as the strongest sub-1M-parameter on-device OV-KWS baselines. We re-implement both per their original specifications to ensure fair, controlled comparisons; our LibriPhrase/GSC numbers closely match reported results (\autoref{fig:results-pack}). 

We train all models with Adam (2500 warm-up steps), batch size 1024, for 50k
steps. Phonemes are encoded using the CMU dictionary (73 tokens). Models are
implemented in PyTorch on x86 Linux with four RTX 4090 GPUs. 


\section{Results}
\label{sec:results}

\footnotetext[1]{We used the official pytorch implementation of \cite{lee2023phonmatchnet} from their GitHub repository \url{https://github.com/ncsoft/PhonMatchNet}.}
\footnotetext[2]{We implemented the model based on configurations from the original paper.}







\subsection{Partial overlap is a major failure mode in baseline models}
When training with only the LibriPhrase dataset, both baselines are strong on prior benchmarks yet fail under partial overlap conditions. SLiCK shows weak performance on POB-Spark, yielding an EER of 64.4\% and an AUC of 31.3\%. Similarly, PhonMatchNet's performance is also limited, with an EER of 35.0\% and an AUC of 70.1\%. This illustrates that partial-overlap phrases are a primary failure mode when models lack explicit overlap exposure or deliberate architectural considerations.

\subsection{EPS reduces prefix bias with minimal in-domain change}
Replacing the scoring layer with EPS markedly improves overlap robustness while keeping performance on prior metrics nearly unchanged. Compared to SLiCK, SLiCK-EPS reduces EER on POB-Spark dataset by 35.1\%, dropping from 64.4\% to 29.3\%, while improving accuracy (ACC) by 46.2\%, from 31.3\% to 77.5\%. It also improves accuracy on POB-LP from 87.6\% to 96.8\%. Compared to PhonMatchNet, SLiCK-EPS reduces POB-Spark EER from 35.0\% to 29.3\% and improves ACC in POB-LP from 64.3\% to 96.8\%. These performance gains on the POB benchmark come without sacrificing previous metrics, with SLiCK-EPS achieving a slight improvement over both baseline models. 

\subsection{POB augmentation improves POB performance but degrades prior metrics}

Adding POB data further improves the overlap robustness for all models. \emph{SLiCK-EPS + POB training} achieves the best performance on POB-Spark regarding both EER (16.2\%) and AUC (91.1\%). However, we observe a performance degradation for all models on the original LibriPhrase and GSC metrics. This suggests that naively adding POB-training data in the current manner may not be the best approach. More work is required to examine the optimal way to incorporate data outside the domain.

\subsection{Best cross-domain balance comes from combining EPS and POB}
As mentioned previously, adding POB data during training improves POB benchmark metrics at the expense of the original metrics. However, \emph{SLiCK-EPS + POB training} introduces the lowest degradation while delivering the strongest performance in POB-Spark. The performance regression on GSC -- which consists of only 1-word utterances, suggests a data composition conflict. The longer-overlap priors introduced by POB datasets may lead the model to de-emphasize the limited phonetic information in brief, single-word utterances. This indicates that while EPS improves robustness for longer, prefix-sharing commands, short-phrase sensitivity requires more nuanced data balancing and weight regularization design in future work.

\section{Conclusion}

Open-vocabulary keyword spotting enables customized, on-device voice control, yet existing models perform poorly in conditions where \emph{negative} enrollment-query pairs share a common prefix, a case underrepresented in standard benchmarks. To make this failure mode measurable, we introduce Partial Overlap Benchmark (POB) consisting of both real and synthesized data; and a lightweight Equal-weighting Position Scoring (EPS) layer that reduces positional bias at the final scoring stage of OV-KWS systems. Evaluations against sub-1M-parameter baselines show that EPS alone restores much of the lost robustness in partial overlap conditions while preserving performance on original benchmark metrics. With POB data added in training, SLiCK-EPS achieves the strongest POB results and the smallest regressions on LibriPhrase and GSC dataset metrics compared to baselines. While adding POB data during training further improves results, it introduces a non-trivial performance trade-off on short commands like GSC. We surface this tension as a direction for future work, focusing on optimized data composition to better balance accuracy across diverse phrase lengths and evaluation regimes.




\vfill\pagebreak\clearpage
\bibliographystyle{IEEEbib}
\bibliography{refs}

\end{document}